\documentstyle[12pt]{article}
\newcommand{\sslash}{\not \!\! s}

\begin{document}

\begin{flushright}{UT-982}
\end{flushright}
\vskip 0.5 truecm

\begin{center}
{\Large{\bf Lattice chiral symmetry, Yukawa couplings and 
the Majorana condition
 }}
\end{center}
\vskip .5 truecm
\centerline{\bf Kazuo Fujikawa and Masato Ishibashi}
\vskip .4 truecm
\centerline {\it Department of Physics,University of Tokyo}
\centerline {\it Bunkyo-ku,Tokyo 113,Japan}
\vskip 0.5 truecm

\makeatletter
\@addtoreset{equation}{section}
\def\theequation{\thesection.\arabic{equation}}
\makeatother

\begin{abstract}
It is shown that the conflict between lattice chiral symmetry
and the Majorana condition in the presence of Yukawa couplings,
which was noted in our previous paper, is  related in an 
essential way to the basic properties of Ginsparg-Wilson 
operators, namely, locality and species doubling. 

\end{abstract}

\section{\bf Introduction}

Recent developments in the treatment of lattice fermions
paved a way to deal with lattice chiral symmetry in a unified 
manner\cite{ginsparg}-\cite{niedermayer}. 
In a recent paper, we pointed out that 
the otherwise successful lattice chiral symmetry has a certain
conflict with the definition of the Majorana fermion 
in the presence of Yukawa couplings\cite{fuji-ishi}. This issue 
was discussed in connection with the lattice regularization of 
the simplest supersymmetric theory\cite{dondi}-\cite{bietenholz}
, namely, the Wess-Zumino model and its non-renormalization 
theorem\cite{wess}-\cite{fujikawa}. 
We consider that a consistent 
formulation of Majorana  fermions is a prerequisite 
for a precise analysis of supersymmetry and its breaking, if 
one adopts the Ginsparg-Wilson operator as a basic building 
block.

In this letter, we further clarify this conflict of lattice
chiral symmetry and the Majorana condition in the presence of 
Yukawa couplings. Our basic observation
is the transformation of a chiral symmetric lattice
theory with Yukawa couplings, which is defined in the manner of 
Niedermayer\cite{niedermayer}, Narayanan\cite{narayanan} and 
Chandrasekharan\cite{chandrasekharan}, to a theory 
which is a generalization of the  model noted by
L\"{u}scher\cite{luscher} by a singular field re-definition. By 
this way we can
understand the origin of the conflict between the 
lattice chiral symmetry and the Majorana condition from a 
different view point. Our 
analysis indicates that the above conflict is related in
an  essential and subtle way to the basic issues of lattice 
chiral symmetry, namely, locality and species doubling. 

In our analysis, we use a  general class of Ginsparg-Wilson 
 operators and our analysis below is valid for all these
operators. The lattice Dirac operator $D$ is defined by the 
algebraic relation\cite{fujikawa2} 
\begin{equation}
\gamma_{5}(\gamma_{5}D)+(\gamma_{5}D)\gamma_{5}=
2a^{2k+1}(\gamma_{5}D)^{2k+2}
\end{equation}
where the parameter $a$ is the lattice spacing; $k$ stands for 
non-negative integers, and $k=0$ 
corresponds to the conventional Ginsparg-Wilson 
relation\cite{niedermayer}.  
When one defines a hermitian operator $H$ by
\begin{equation}
H=a\gamma_{5}D=H^{\dagger}=aD^{\dagger}\gamma_{5}
\end{equation}
the above algebraic relation is written as
\begin{equation}
\gamma_{5}H+H\gamma_{5}=2H^{2k+2}.
\end{equation}
We can also show
\begin{equation}
\gamma_{5}H^{2}=(\gamma_{5}H+H\gamma_{5})H-
H(\gamma_{5}H+H\gamma_{5})+ H^{2}\gamma_{5}=H^{2}\gamma_{5}
\end{equation}
which implies
\begin{equation}
H^{2}=a^{2}D^{\dagger}D=\gamma_{5}H^{2}\gamma_{5}
=a^{2}DD^{\dagger}.
\end{equation}
When we define
\begin{eqnarray}
&&\Gamma_{5}\equiv \gamma_{5}-H^{2k+1},\nonumber\\
&&\hat{\gamma}_{5}\equiv \gamma_{5}-2H^{2k+1},
\end{eqnarray}
the defining algebra (1.1) is written as
\begin{equation}
\gamma_{5}H+H\hat{\gamma}_{5}=0
\end{equation}
or $\Gamma_{5}H+H\Gamma_{5}=0$, and $(\hat{\gamma}_{5})^{2}=1$.
We can also show the relation
\begin{equation}
\gamma_{5}\Gamma_{5}+\Gamma_{5}\gamma_{5}=2\Gamma_{5}^{2}
=2(1-H^{4k+2})
\end{equation}
which implies $H^{2}\leq 1$.
We next note\cite{niedermayer}   
\begin{equation}
D = P_{+}D\hat{P}_{-} +  P_{-} D\hat{P}_{+}.    
\end{equation}
Here we defined two projection operators
\begin{eqnarray}
P_{\pm} &=& \frac{1}{2}( 1 \pm \gamma_{5}),\nonumber\\
\hat{P}_{\pm} &=& \frac{1}{2}( 1 \pm \hat{\gamma}_{5})
\end{eqnarray}
which satisfy the relations
\begin{eqnarray}
&&P_{+}\hat{P}_{+}=P_{+}\gamma_{5}\Gamma_{5},
\nonumber\\
&&P_{-}\hat{P}_{-}=P_{-}\gamma_{5}\Gamma_{5}.
\end{eqnarray}
We then define the chiral 
components\cite{niedermayer}\cite{narayanan}
\begin{equation}
\bar{\psi}_{L,R}= \bar{\psi}P_{\pm}, \ \ 
\psi_{R,L}= \hat{P}_{\pm}\psi
\end{equation}
and the scalar and pseudscalar densities 
by\cite{chandrasekharan} 
\begin{eqnarray}
S(x) &=& \bar{\psi}_{L}\psi_{R} + \bar{\psi}_{R}\psi_{L}
=\bar{\psi}\gamma_{5}\Gamma_{5}\psi,
\nonumber\\
P(x) &=& \bar{\psi}_{L}\psi_{R} - \bar{\psi}_{R}\psi_{L}
=\bar{\psi}\Gamma_{5}\psi.
\end{eqnarray}

\section{Yukawa couplings and the Majorana condition}

The most natural Lagrangian 
consistent with lattice chiral symmetry 
$\delta\psi=i\epsilon\hat{\gamma}_{5}\psi$, 
$\delta\bar{\psi}=\bar{\psi}i\epsilon\gamma_{5}$ and 
$\delta\phi=-2i\epsilon\phi$, which is 
softly broken by the mass term, is defined by\cite{niedermayer}
\cite{chandrasekharan}
\begin{eqnarray}
{\cal L}&=&\bar{\psi}D\psi 
+ m\bar{\psi}\gamma_{5}\Gamma_{5}\psi
+ 2g\bar{\psi}(P_{+}\phi\hat{P}_{+}
+P_{-}\phi^{\dagger}\hat{P}_{-})\psi\nonumber\\
&=&\bar{\psi}_{R}D\psi_{R}+\bar{\psi}_{L}D\psi_{L} 
+ m[\bar{\psi}_{R}\psi_{L}+\bar{\psi}_{L}\psi_{R}]\nonumber\\
&&+ 2g[\bar{\psi}_{L}\phi\psi_{R}
+\bar{\psi}_{R}\phi^{\dagger}\psi_{L}].  
\end{eqnarray}
We fixed the mass term in such a way that it is 
generated by a shift 
$\phi(x)\rightarrow\phi(x)+m/(2g)$ in 
$\phi(x)=(A(x)+iB(x))/\sqrt{2}$ in the interaction terms; we 
adopt this procedure in the following.
The fermion mass term is then defined by the scalar density 
formed of a fermion bi-linear (1.13). 

It has been shown elsewhere\cite{fuji-ishi} that
the above Lagrangian (2.1) has a difficulty in performing the 
Majorana reduction, and thus the Majorana fermion is not 
defined in a manner consistent with lattice chiral symmetry: 
When  one 
defines\cite{suzuki}\cite{nicolai}\cite{van nieuwenhuizen} 
\begin{eqnarray}
&&\psi=(\chi+i\eta)/\sqrt{2},\nonumber\\
&&\bar{\psi}=(\chi^{T}C-i\eta^{T}C)/\sqrt{2}
\end{eqnarray}
in ${\cal L}$ (2.1), one naively expects
\footnote{If $(CO)^{T}=-CO$ 
for a general operator $O$, the cross
term vanishes $\eta^{T}CO\chi-\chi^{T}CO\eta=0$ by using the 
anti-commuting property of $\chi$ and $\eta$.}
by noting $P_{+}\phi\hat{P}_{+}
+P_{-}\phi^{\dagger}\hat{P}_{-}=\frac{1}{\sqrt{2}}
(A\gamma_{5}\Gamma_{5}+iB\Gamma_{5})$,
\begin{eqnarray}
{\cal L}&=&\frac{1}{2}\chi^{T}CD\chi 
+ \frac{1}{2}m\chi^{T}C\gamma_{5}\Gamma_{5}\chi
+ g\frac{1}{\sqrt{2}}\chi^{T}C(A\gamma_{5}\Gamma_{5}
+iB\Gamma_{5})\chi\nonumber\\
&+&\frac{1}{2}\eta^{T}CD\eta 
+ \frac{1}{2}m\eta^{T}C\gamma_{5}\Gamma_{5}\eta
+ g\frac{1}{\sqrt{2}}
\eta^{T}C(A\gamma_{5}\Gamma_{5}+iB\Gamma_{5})\eta 
\end{eqnarray}
but this actually fails, since   
$(C\Gamma_{5})^{T}\neq -C\Gamma_{5}$, where $C$ stands for the 
charge conjugation matrix, and the non-commuting property of the 
difference operator $[\gamma_{5}\Gamma_{5},A(x)]\neq 0$ 
though $(C\gamma_{5}\Gamma_{5})^{T}= -C\gamma_{5}\Gamma_{5}$
 or equivalently 
$C\gamma_{5}\Gamma_{5}C^{-1}=(\gamma_{5}\Gamma_{5})^{T}$ in 
the Yukawa couplings. To be precise, we need to perform a charge 
conjugation operation of the gauge field to satisfy
$C\gamma_{5}\Gamma_{5}C^{-1}=(\gamma_{5}\Gamma_{5})^{T}$, for 
example, in the presence of the background gauge field; this 
extra operation of charge conjugation is implicitly assumed in 
the following.

We now observe that the  field 
re-definition\footnote{A related transformation has been 
discussed in the past in a different context, namely, to 
relate the domain-wall fermion to the overlap 
fermion\cite{neuberger3}.}
\begin{eqnarray}
&&\psi^{\prime}=\gamma_{5}\Gamma_{5}\psi,\nonumber\\
&&\bar{\psi}^{\prime}=\bar{\psi}
\end{eqnarray}
in the above Lagrangian gives rise to the Lagrangian
\begin{eqnarray}
{\cal L}^{\prime}&=&\bar{\psi}^{\prime}D
\frac{1}{\gamma_{5}\Gamma_{5}}\psi^{\prime} 
+ m\bar{\psi}^{\prime}\psi^{\prime}
+ 2g\bar{\psi}^{\prime}(P_{+}\phi P_{+}
+P_{-}\phi^{\dagger}P_{-})\psi^{\prime} 
\end{eqnarray}
where we used the relations (1.11). 
This shows that the theory defined by the Lagrangian
invariant under the lattice chiral symmetry
\begin{equation}
Z=\int{\cal D}\psi{\cal D}\bar{\psi}\exp[\int {\cal L}d^{4}x ]
\end{equation}
is related to the theory defined by the transformed Lagrangian
as 
\begin{equation}
Z=(\det \gamma_{5}\Gamma_{5})Z^{\prime}\equiv 
(\det \gamma_{5}\Gamma_{5})\int{\cal D}\psi^{\prime}
{\cal D}\bar{\psi}^{\prime}\exp[\int {\cal L}^{\prime}d^{4}x].
\end{equation} 

This new Lagrangian (2.5) 
corresponds to a generalization of the Lagrangian considered 
by L\"{u}scher if one eliminates the auxiliary field. To be 
specific, L\"{u}scher considered the Lagrangian\cite{luscher}
\begin{eqnarray}
{\cal L}_{L}&=&\bar{\psi}D\psi- \frac{1}{a}\bar{\chi}\chi 
+2g(\bar{\psi}+\bar{\chi})(P_{+}\phi P_{+}
+P_{-}\phi^{\dagger}P_{-})(\psi +\chi)
\end{eqnarray}
which is shown to be invariant under a modified chiral
 transformation, if one assumes that $D$ satisfies the 
Ginsparg-Wilson relation (1.1) with $k=0$, namely, the overlap 
operator\cite{neuberger}.
By considering  the re-definition of field variables 
\begin{eqnarray}
&&\psi^{\prime}=(\psi +\chi),\ \ \ \
\bar{\psi}^{\prime}=(\bar{\psi}+\bar{\chi}),
\nonumber\\ 
&&\Psi=(\psi -\chi),\ \ \ \  
\bar{\Psi}=(\bar{\psi}-\bar{\chi})
\end{eqnarray}
one obtains\footnote{This was discussed  in a different 
context in Ref.\cite{chiu2}.} after the path integral over 
$\Psi$ and
$\bar{\Psi}$
\begin{equation}
\int{\cal D}\bar{\psi}{\cal D}\psi{\cal D}\bar{\chi}{\cal D}
\chi\exp[\int d^{4}x{\cal L}_{L}]
=\det(1-aD)\int{\cal D}\bar{\psi}^{\prime}{\cal D}\psi^{\prime}
\exp[\int d^{4}x{\cal L}^{\prime}_{L}]
\end{equation}
where 
\begin{equation}
{\cal L}^{\prime}_{L}
=\bar{\psi}^{\prime}D\frac{1}{(1-aD)}\psi^{\prime}  
+ 2g\bar{\psi}^{\prime}(P_{+}\phi P_{+}
+P_{-}\phi^{\dagger}P_{-})\psi^{\prime}.
\end{equation}
This 
final expression (2.11) corresponds to our Lagrangian (2.5)
for $k=0$ (i.e., the standard overlap 
operator for which $\gamma_{5}\Gamma_{5}=1-aD$), up to the 
chiral symmetry breaking mass term.

The above transformation (2.4) 
is singular if $\gamma_{5}\Gamma_{5}=1-\gamma_{5}HH^{2k}=0$.
Since $\Gamma^{2}_{5}=1-H^{4k+2}$, the necessary condition for
the appearance of the singularity is $H^{2}=1$.
This condition is analyzed in more detail as follows:
$H^2$ is found from an explicit form 
of $H$, which is local and free of 
species doubling\cite{fujikawa3}\cite{chiu},
\begin{eqnarray}
H(ap_{\mu})&=&\gamma_{5}(\frac{1}{2})^{\frac{k+1}{2k+1}}
(\frac{1}{\sqrt{H^{2}_{W}}})^{\frac{k+1}{2k+1}}
\{(\sqrt{H^{2}_{W}}+M_{k})^{\frac{k+1}{2k+1}}
-(\sqrt{H^{2}_{W}}-M_{k})^{\frac{k}{2k+1}}
\frac{\sslash}{a} \},\nonumber
\end{eqnarray}
where $\sslash=\gamma^{\mu}\sin ap_\mu$ with anti-hermitian 
$\gamma^{\mu}$ and
\begin{eqnarray}
M_k(p)&=&\left(\frac{r}{a}\sum_{\mu}(1-\cos ap_\mu)
\right)^{2k+1}-\left(\frac{m_0}{a}\right)^{2k+1},
\nonumber\\
H_W^2&=&\left(\sum_{\mu}\frac{1}{a^2}\sin^2 ap_\mu
\right)^{2k+1}+\left(M_k(p)\right)^2.
\end{eqnarray}
We thus obtain 
\begin{eqnarray}
H^2(ap)&=&(\frac{1}{2\sqrt{H^{2}_{W}}})^{\frac{2k+2}{2k+1}}
\{(\sqrt{H^{2}_{W}}+M_{k})^{\frac{2k+2}{2k+1}}
+(\sqrt{H^{2}_{W}}-M_{k})^{\frac{2k}{2k+1}}
\frac{s^2}{a^2} \}.\nonumber\\
\end{eqnarray}
Using the relation 
$s^2/a^2=(H_W^2-M_k^2)^{\frac{1}{2k+1}}$,
we find that $H^2(ap)=1$ implies $\sqrt{H^{2}_{W}}=M_k.$
This  last condition is written explicitly  as
\begin{eqnarray}
&&\sqrt{\left(\sum_{\mu}\frac{1}{a^2}\sin^2 ap_\mu
\right)^{2k+1}+\left(M_k(p)\right)^2}=M_k(p).
\end{eqnarray}
Since $m_0$ is constrained by $0<m_0<2r$ to avoid the 
appearance of species doublers, we have 
$M_k<0$ for
a physical mode ( $p_\mu=(0,0,0,0)$ ) and $M_k>0$ for doubler
modes ( $p_\mu=(\pi/a,0,0,0)$, etc ). Therefore the above
equation (2.14) holds only for would-be doubler modes.
Just on top of the doubler modes, one can also confirm  
\begin{equation}
\gamma_{5}\Gamma_{5}=1-\gamma_{5}HH^{2k}=0.
\end{equation}
Thus the  singularity
of the transformation comes from the momentum regions of 
would-be species doublers.
 It is interesting to notice that when $m_0>8r$ 
(where all the doubler modes appear as massless modes), the 
singularity of the above transformation disappears.

The above transformation (2.4) thus induces 
singularities inside the Brillouin zone and thus spoils the 
locality of the Dirac operator $D^{\prime}\equiv D/(\gamma_{5}
\Gamma_{5})$. The ordinary formulation (2.1), which is local, 
and the model 
(2.8), which appears to be local but actually non-local, 
formally give rise to the same path integeral as in (2.7) and 
(2.10), but this equivalence does not hold in a strict sense 
since one has to go through the {\em singular Lagrangian} such 
as (2.5) in the intermediate stage.

The interesting property of this field re-definition (2.4) 
in the context of the present analysis of the Majorana condition
is that the transformed singular theory (2.5) is invariant under 
the naive continuum chiral symmetry and that it allows the 
Majorana reduction. This fact is understood
by noting the relation
\begin{eqnarray}
\gamma_{5}\Gamma_{5}\hat{\gamma}_{5}
=\gamma_{5}(\gamma_{5}\Gamma_{5}+\Gamma_{5}\gamma_{5})
-\gamma_{5}\Gamma_{5}\gamma_{5}
=\gamma_{5}(\gamma_{5}\Gamma_{5})
\end{eqnarray}
where we used (1.8).
This relation shows that the chiral transformation of $\psi$ 
generated by $\hat{\gamma}_{5}$ of the regular theory
is related to the chiral transformation of $\psi^{\prime}$
generated by the continuum $\gamma_{5}$ as
\begin{eqnarray}
\delta\psi^{\prime}&\equiv&(\gamma_{5}\Gamma_{5})\delta\psi
=\gamma_{5}\Gamma_{5}i\epsilon\hat{\gamma}_{5}\psi\nonumber\\
&=&i\epsilon\gamma_{5}(\gamma_{5}\Gamma_{5})\psi
=i\epsilon\gamma_{5}\psi^{\prime}
\end{eqnarray}
and of course $
\delta\bar{\psi}^{\prime}=\delta\bar{\psi}
=\bar{\psi}^{\prime}i\epsilon\gamma_{5}$.
As for an analysis of the Majorana reduction
of this transformed singular Lagrangian, we 
note that 
\begin{eqnarray}
(CD\frac{1}{\gamma_{5}\Gamma_{5}})^{T}&=&
(CD\frac{1}{C\gamma_{5}\Gamma_{5}}C)^{T}
=C^{T}(\frac{1}{C\gamma_{5}\Gamma_{5}})^{T}(CD)^{T}\nonumber\\
&=&-C\frac{1}{C\gamma_{5}\Gamma_{5}}CD
=-C\frac{1}{\gamma_{5}\Gamma_{5}}D
=-CD\frac{1}{\gamma_{5}\Gamma_{5}}
\end{eqnarray}
where $C$ is the charge conjugation matrix and we used the 
properties\cite{fuji-ishi}\cite{suzuki}
\begin{eqnarray}
&&C^{T}=-C,\nonumber\\
&&(C\gamma_{5}\Gamma_{5})^{T}=-C\gamma_{5}\Gamma_{5},\nonumber\\
&&(CD)^{T}=-CD
\end{eqnarray}
and the relation $
D\gamma_{5}\Gamma_{5}=D(1-\gamma_{5}HH^{2k})=
\gamma_{5}\Gamma_{5}D$
by noting  $DH^{2}=H^{2}D$ which follows from (1.5).
Our operator thus satisfies the condition necessary for the 
Majorana reduction, if one ignores the singularity in 
$CD/(\gamma_{5}\Gamma_{5})$.

When  one performs the field transformation (2.2) ( with 
$\psi$ replaced by $\psi^{\prime}$) 
in ${\cal L}^{\prime}$ (2.5) by noting
$(C\gamma_{5})^{T}=-C\gamma_{5}$,
one can thus define the Majorana fermion in a formal sense by
\begin{eqnarray}
{\cal L}^{\prime}_{Majorana}
&=&\frac{1}{2}\chi^{T}CD\frac{1}{\gamma_{5}\Gamma_{5}}\chi 
+ \frac{1}{2}m\chi^{T}C\chi
+ g\chi^{T}C(P_{+}\phi P_{+}
+P_{-}\phi^{\dagger}P_{-})\chi.  
\end{eqnarray}
This formulation of the Majorana fermion and the resulting 
Pfaffian, if one does not care 
about the singularity, gives rise to the same result 
as in our previous paper\cite{fuji-ishi} which utilized 
$\sqrt{Z}$, on the basis of the relation 
\begin{equation}
\sqrt{(\det \gamma_{5}\Gamma_{5})Z^{\prime}}=\sqrt{Z}
\end{equation}
in a formal perturbation theory, for example\footnote{ In the 
non-perturbative sense, the relation (2.21) stands for 
something like $0\times(1/0)=1$ if one takes possible zero 
modes in $\gamma_{5}\Gamma_{5}$ into account.}.

If one tentatively adopts the singular Lagrangian (2.20), a 
lattice version of the  Wess-Zumino model in our 
previous paper\cite{fuji-ishi} is re-written as (after a 
rescaling of the auxiliary field $F$) 
\begin{eqnarray}
{\cal L}_{WZ}&=&\frac{1}{2}\chi^{T}C\frac{1}{\Gamma_{5}}
\gamma_{5}D\chi + \frac{1}{2}m\chi^{T}C\chi
+ g\chi^{T}C(P_{+}\phi P_{+}+P_{-}\phi^{\dagger}P_{-})\chi
\nonumber\\
&&-\phi^{\dagger}D^{\dagger}D\phi
+F^{\dagger}\frac{1}{\Gamma^{2}_{5}}F
+m[F\phi+(F\phi)^{\dagger}]
+g[F\phi^{2}+(F\phi^{2})^{\dagger}].
\end{eqnarray}
Note that $((1/\Gamma_{5})\gamma_{5}D)^{2}=-D^{\dagger}D
(1/\Gamma^{2}_{5})$, and thus the kinetic (K\"{a}hler) terms 
satisfy a 
necessary condition for supersymmetry provided that one ignores
the $4\times 4$ unit matrix in $D^{\dagger}D$ and 
$1/\Gamma^{2}_{5}$ in bosonic terms. The 
(super-)potential parts of this Lagrangian (2.22) are identical 
to those of the continuum theory. This representation of the 
Lagrangian, when treated with due qualifications,  is thus 
useful to understand the symmetry aspects of the model. 
But the formulation of this Lagrangian is not satisfactory 
since  it does not give a uniform wave function renormalization 
factor in the one-loop level of perturbation
theory\cite{fuji-ishi}, besides the issues related to the 
Leibniz rule\cite{dondi}. 
\\

Alternatively, one may consider the symmetric  
definitions\cite{hasenfratz2} of 
left and right components by $\psi_{R,L}
=(1\pm \Gamma_{5}/\Gamma)\psi/2$ and 
$\bar{\psi}_{R,L}=\bar{\psi}(1\mp \gamma_{5}\Gamma_{5}\gamma_{5}
/\Gamma)/2$, where $\Gamma=\sqrt{1-H^{4k+2}}$ 
, respectively. In this case, the second 
expression of (2.1) is written as 
\begin{eqnarray}
{\cal L}&=&\bar{\psi}D\psi 
+ m\bar{\psi}\gamma_{5}\Gamma_{5}\psi
+ \frac{g}{\sqrt{2}}\bar{\psi}[A+
(\gamma_{5}\Gamma_{5}\gamma_{5}/\Gamma)A
(\Gamma_{5}/\Gamma)
+i(\gamma_{5}\Gamma_{5}\gamma_{5}/\Gamma)B
+iB(\Gamma_{5}/\Gamma)]\psi\nonumber\\
\end{eqnarray}
 and the Majorana fermion can be defined. But the modified 
chiral operators, $\Gamma_{5}/\Gamma$ and 
$\gamma_{5}\Gamma_{5}\gamma_{5}/\Gamma$, are ill-defined at 
$H^{2}=1$, namely
\begin{equation}
\Gamma_{5}/\Gamma\simeq \gamma_{5}(\gamma^{\mu}\sin ap_\mu/a)
/\sqrt{\sum_{\mu}\sin^2 ap_\mu/a^2}
\end{equation}
for $H^{2}\simeq 1$ by noting (2.14), and Yukawa couplings 
become  singular.  
Incidentally, an analysis of the Euclidean Majorana condition is
 related to that of CP 
symmetry\cite{fuji-ishi}\cite{hasenfratz2}.
\\

We summarize our analysis as follows:\\
1. The most natural formulation consistent with lattice
chiral symmetry (2.1), which is successful in QCD, does not
accommodate the Euclidean Majorana fermion\cite{fuji-ishi}. \\
2. If one allows a non-local singular Lagrangian such as (2.5), 
(2.11) 
and (2.23) (or if one allows species doubling by choosing 
$m_0>8r$), one can accommodate the Euclidean Majorana fermion 
and the resulting Pfaffian. The non-locality is however expected
 to become serious in the presence of background gauge 
field\cite{chiu2}.\\
3. Our analysis is valid for a general class of Ginsparg-Wilson
operators (1.1) and thus exhibits generic properties of these 
lattice Dirac operators.
\\

In conclusion, a deeper understanding of Ginsparg-Wilson 
operators is required to incorporate Majorana fermions with 
Yukawa couplings in a manner consistent with lattice chiral
symmetry.

\end{document}